\documentclass[aps,prd,preprintnumbers,showpacs]{revtex4}
\usepackage[dvips]{graphicx}
\begin{document}

%
%

\eprint{Nisho-1-2013}
\title{Rapid Decay of Color Gauge Fields by Production of Magnetic Monopoles}
\author{Aiichi Iwazaki}
\affiliation{International Economics and Politics, Nishogakusha University,\\ 
6-16 3-bantyo Chiyoda Tokyo 102-8336, Japan.}   
\date{Aug. 18, 2013}
\begin{abstract}
It has been argued that the rapid decay of coherent color gauge fields
generated immediately after high energy heavy ion collisions
produces quark gluon plasma. But there are no convincing mechanisms
for the rapid decay of the gauge fields which satisfy phenomenological
constraints on their life time.
We show by using classical statistical field theory that the production of 
magnetic monopoles cause the rapid decay of the gauge fields.
The monopoles are unstable modes just like Nielsen-Olesen modes
and are enormously produced by dual Schwinger mechanism. 
\end{abstract}
\hspace*{0.3cm}
\hspace*{1cm}

\maketitle

\section{introduction}

One of the most significant unsolved puzzles in the dynamics of 
quarks and gluons in high energy heavy ion collisions
is how coherent color gauge fields\cite{cgc} ( here we call them glasma ) 
generated immediately after the collisions decay into quark gluon plasma (QGP). 
In particular, the fields rapidly decay and produce thermalized QGP
within a time less than $1$fm/c\cite{hirano}.
The puzzle is what kind of mechanism causes such a rapid decay of the glasma.

The presence of such coherent gauge fields has been
shown using a model of color glass condensate\cite{cgc}.
They are color electric and magnetic fields pointed
in longitudinal direction. 
They are uniform in the longitudinal direction, while they
vary in transverse directions. Thus, we may think that they have
the form of flux tubes with various widths. 
The typical width is
given by $Q_s^{-1}$; $Q_s$ is saturation momentum in the collisions.
Similarly, the typical values of the gauge field strength are 
given by $Q_s^2$.

Instabilities of such gauge fields have been shown 
in numerical simulations\cite{berges,berges1,kunihiro,ven,fuku}. 
It has been found that small 
fluctuations added to the gauge fields grow exponentially.
Consequently, the longitudinal pressure of gauge fields or the distance between
classical trajectories in space of gauge fields grow exponentially. 
The presence of the instabilities implies
 that the gauge fields decay with entropy production\cite{kunihiro}
and indicates that they eventually produce
QGP. However,
the life time of the gauge field caused by the instabilities
is much longer than
$1$fm/c. 


The instabilities have been shown 
to be Nielsen-Olesen instability\cite{nielsen,instability}. Actually, 
the authors of the paper\cite{berges2}
has demonstrated numerically in detail that
they are Nielsen-Olesen instabilities, not Weibel instabilities.
Nielsen and Olesen have shown in their original paper\cite{nielsen}
that there exist unstable modes growing exponentially with time $\sim \exp(\gamma t)$ 
under homogeneous color magnetic field $B$.   
Their growth rate $\gamma$ is given 
by $\sqrt{gB}$; $g \,\,( >0 )$ is a gauge coupling constant.
Since the magnetic fields in the glasma are inhomogeneous,
unstable modes grow more slowly than in the case of the homogeneous magnetic field.
Indeed, $\gamma$ is given by not square root $\sqrt{gB}\sim Q_s$ of
magnetic fields,
but square root $\sqrt{gB_{\rm{eff}}} $ of effective magnetic field $gB_{\rm{eff}}$,
which is much smaller than $Q_s^2$, e.g. $\sqrt{gB_{\rm{eff}}}\sim 0.2Q_s$. 
That is, because $gB$ varies spatially and temporally, 
the average growth rate of the unstable modes is much small compared with that in the case of 
homogeneous magnetic fields. 
Furthermore, the growth rates\cite{ven,fuku, berges1} in expanding glasma are much smaller than
those in non-expanding glasma, because background magnetic fields become weak 
with the expansion\cite{lap}.
Therefore, 
the life time of the glasma caused by the instabilities 
is longer than 
$1$fm/c. 
This contradicts
the phenomenological analysis,
which shows that thermalized QGP is realized within a time less than $1$fm/c.
 
Magnetic monopoles\cite{coleman} are essential ingredients for
quark confinement. Their condensation realizes a confining vacuum of dual superconductor\cite{dual}.
They have ``imaginary mass" so that
the fields representing monopoles exponentially increase
just like Nielsen-Olesen unstable modes. 
The mass is homogeneous and 
its absolute value is of the order of
$1$GeV\cite{dual, koma}. Furthermore, magnetic charge $g_m=2\pi/g$ is large
for small gauge coupling $\alpha_s\equiv 4\pi/g^2$ in high energy heavy ion collisions.  
Hence, we may expect that the monopoles are enormously produced in the collisions
and their production causes a rapid decay of the glasma.  
 
\vspace{0.3cm}

In this paper we show that
the glasma rapidly decays by the production of the magnetic monopoles.
The production arises by the Schwinger mechanism and its back reaction
to background color magnetic fields causes the fields decay. 
The calculation is performed by using
classical statistical field theory\cite{csft, csft1, csft2}. Thus, we take into account 
one loop quantum effects of monopole fields in the calculation.
We find that the life time of the color magnetic fields is much shorter than $1$fm/c. 
For comparison, we also calculate the back reaction of the Nielsen-Olesen unstable modes
by using classical statistical field theory.
We find that the glasma decays ten times faster by the production of the monopoles than
by the Nielsen-Olesen unstable modes.   
Although we only treat non-expanding glasma,
our result does hold even in expanding glasma.

In next section \ref{2}, we explain the presence of Nielsen-Olesen instabilities under 
inhomogeneous magnetic fields. In section \ref{3}, after discussing 
monopole production by dual Schwinger mechanism, we show that magnetic fields
rapidly decay by the monopole production. Discussions and conclusions follow
in section \ref{4}


\section{Nielsen-Olesen instability in inhomogeneous magnetic fields}
\label{2}

First, we briefly review the Nielsen-Olesen unstable modes
found in the previous numerical
simulations. 
In particular, we show that they are present in any inhomogeneous
background magnetic fields and argue
why the growth rates of the unstable modes are
much smaller in inhomogeneous magnetic fields
than those in homogeneous magnetic fields.
Although the glasma involves the inhomogeneous magnetic fields,
we can describe the properties of 
the instabilities by using effective homogeneous magnetic fields.

We consider SU(2) gauge theory with
the background color electric and magnetic fields given by
$\vec{E}_a=\delta_{a,3}\vec{E}$ and $\vec{B}_a=\delta_{a,3}\vec{B}$.
Here we assume for simplicity that these fields are pointed into $3$ direction of color space.
The fields are represented by the gauge potential $A_{\mu}\equiv A_{\mu}^{a=3}$.
Under the background fields, the gauge potentials $\Phi_{\mu}\equiv (A_{\mu}^1+iA_{\mu}^2)/\sqrt{2}$ 
perpendicular to $A_{\mu}^3$ behave
as charged vector fields. When we represent SU(2) gauge potentials $A_{\mu}^a$ using 
the variables $A_{\mu}$ and $\Phi_{\mu}$, Lagrangian of SU(2) gauge potentials is
written in the following,

\begin{equation}
\label{L}
L=-\frac{1}{4}F_{\mu,\nu}^2-\frac{1}{2}|D_{\mu}\Phi_{\nu}-D_{\nu}\Phi_{\mu}|^2
-ig(\partial_{\mu}A_{\nu}-\partial_{\nu}A_{\mu})\Phi^{\mu}\Phi^{\dagger \nu}
+\frac{g^2}{4}(\Phi_{\mu}^{\dagger}\Phi_{\nu}-\Phi_{\nu}^{\dagger}\Phi_{\mu})^2,
\end{equation} 
with $F_{\mu,\nu}=\partial_{\mu}A_{\nu}-\partial_{\nu}A_{\mu}$ and 
$D_{\mu}=\partial_{\mu}+igA_{\mu}$.
We may understand that the fields $\Phi_{\mu}$ represent charged vector fields
with the anomalous magnetic moment described by the term 
$-ig(\partial_{\mu}A_{\nu}-\partial_{\nu}A_{\mu})\Phi^{\mu}\Phi^{\dagger \nu}$. 
This term gives rise to the instability of the background magnetic field.

For example, when a homogeneous background magnetic field 
is given by $B=\partial_1A_2-\partial_2A_1$,
but $E=0$, 
the particles represented by the fields $\Phi_i$
occupy the Landau levels denoted by integer $N\ge 0$. Their energies are 
given by $E_N=\sqrt{2gB(N+1/2)\pm 2gB+p_3^2}$, where $\pm$ denotes magnetic moment
parallel ( $-$ ) or anti-parallel ( $+$ ) to $\vec{B}$
and $p_3$ denotes a momentum component parallel to $\vec{B}$.
Among them the energies of the states 
in the lowest Landau level ( $N=0$ ) 
with the magnetic moment parallel to $\vec{B}$ 
can be imaginary; $E_{N=0}=\sqrt{p_3^2-gB}$ with $p_3^2<gB$. 
Thus, the modes 
with the imaginary energies exponentially increase or decrease with time;
$\Phi_i\propto \exp(-iE_{N=0}t)=\exp(\pm |E_{N=0}|t)=\exp(\pm |\sqrt{gB-p_3^2}|\,t)$.
The modes are called as Nielsen-Olesen
unstable modes. In particular, the mode with $p_3=0$ increases most rapidly
with the growth rate $\sqrt{gB}$.

The ``kinetic energy" ( the second term of the Lagrangian ) of the states in the 
lowest Landau level is 
given by $+gB$ as usual, while the ``potential energy" given by the anomalous magnetic moment
( the third term )
is negative, that is given by $-2gB$.
Thus, the ``total energy" $E^2$ is given such that $gB-2gB=-gB<0$. This leads to  
the imaginary energy $E=\sqrt{-gB}$.

\vspace{0.3cm}

The case of the homogeneous magnetic field is not realistic. 
Magnetic fields in the glasma have the form of flux tubes and oscillate
according to the Maxwell equations.
Then, the ``potential energy" ( $-2gB$ ) 
varies spatially and oscillates, in other words, 
the potential can be negative or positive. 
Although
their typical energy scale is given by the saturation momentum $Q_s$,
the average depth of the potential is not of the order of $Q_s^2$, but much less than $Q_s^2$.
Hence we may ask whether or not states with negative ``total energies" are present even in such cases.
That is, we ask whether or not the instabilities caused by the ``potential" are still present.
We should remember that
the numerical simulations have shown the presence of the instabilities.

We can analytically show the presence of the instability
even in the inhomogeneous magnetic fields.
In order to see it we write down the Hamiltonian of the field $\Phi_i$,

\begin{equation}
H=|\partial_t\Phi_i|^2+\frac{1}{2}|D_i\Phi_j-D_j\Phi_i|^2+igF_{i,j}\Phi_i\Phi_j^{\dagger}
\end{equation}
where we neglected the interaction terms $\sim g^2\Phi_i^4$ and 
used a gauge $\Phi_0=0$. The third term represents the anomalous magnetic moment
of the field $\Phi_i$. 
Obviously, the equations of motion of $\Phi_i$, 
is
given such that 
$-\partial_t^2\Phi_i=\frac{\delta}{\delta\Phi_i^{\dagger}}\int d^3x(
\frac{1}{2}|D_l\Phi_j-D_j\Phi_l|^2+igF_{l,j}\Phi_l\Phi_j^{\dagger})$. 
We can show that for arbitrarily given $\vec{B}$, there are
field configurations $\Phi_i$ such that 
$\int d^3x \big(\frac{1}{2}|D_i\Phi_j-D_j\Phi_i|^2+igF_{i,j}\Phi_i\Phi_j^{\dagger}\big) < 0$.
The fact implies that negative eigenvalues $E^2<0$ 
( $\Phi \propto \exp(iEt)$ ) are present for arbitrarily given $\vec{B}$.

In order to find such field configurations, we put $\Phi_i=D_i\Phi$ and rewrite

\begin{eqnarray}
&\int& d^3x \Big(\frac{1}{2}|D_i\Phi_j-D_j\Phi_i|^2+igF_{i,j}\Phi_i\Phi_j^{\dagger} \Big)
=\int d^3x \Big ( g^2B^2|\Phi|^2
+igF_{i,j}\partial_i\Phi\partial_j\Phi^{\dagger}-g^2B^2|\Phi|^2-g^2\partial_iF_{i,j}A_j|\Phi|^2\Big)= \nonumber \\
&\int& d^3x \Big(-\frac{1}{2}\rm{rot}\vec{B}\cdot(ig\Phi^{\dagger}\vec{D}\Phi
-ig(\vec{D}\Phi)^{\dagger}\Phi)\Big)
=\int d^3x\,\frac{g}{2} \Big(\rm{rot}\vec{B}\cdot(g\vec{A}+\vec{\partial}\theta )|\Phi|^2\Big)
\end{eqnarray}
with $\Phi\equiv|\Phi|\exp(i\theta)$, where we assumed that the field $\Phi$ vanishes at spatial infinity and
used the identity $D_iD_j-D_jD_i=igF_{i,j}$.

In the last equation, we can take an appropriate continuous function $\theta$  such that
the function $\rm{rot}\vec{B}\cdot(g\vec{A}+\vec{\partial}\theta )$ is negative
at a point $P$. Then, from the continuity of the fields there is a region
involving the point $P$ in which $\rm{rot}\vec{B}\cdot(g\vec{A}+\vec{\partial}\theta )$
is still negative. We can take a field configuration $|\Phi|$ such that $|\Phi|$ is sufficiently small
outside the region for the integral  
$\int d^3x\,\frac{g}{2} \Big(\rm{rot}\vec{B}\cdot(g\vec{A}+\vec{\partial}\theta )|\Phi|^2\Big)$
to be negative. Therefore, we find such field configurations $\Phi_i$ that the integral
$\int d^3x \big(\frac{1}{2}|D_i\Phi_j-D_j\Phi_i|^2+igF_{i,j}\Phi_i\Phi_j^{\dagger}\big)$
is negative.
In this way we can see that there are negative eigenvalues $E^2$ 
for arbitrarily given $B$.	
This is the reason why instabilities have been found in the
previous numerical simulations. 

We should note that the integral cannot take arbitrarily large 
negative values because we should take sufficiently small $\Phi_i$ for the quartic interactions
$g^2\Phi_i^4$ to be neglected. Thus, our arguments do not show the presence of
arbitrarily large negative eigenvalues $E^2$.  

Rigorously speaking, in order for the eigenvalues to exist, the Hamiltonian has to be static.
But, because the background magnetic field $B$ varies with time, $H$ depends on time.
What we have shown above is that there exist such eigenstates $\Phi_i$ that 
$E(t)^2\Phi_i=\frac{\delta}{\delta\Phi_i^{\dagger}}\int d^3x(
\frac{1}{2}|D_l\Phi_j-D_j\Phi_l|^2+igF_{l,j}\Phi_l\Phi_j^{\dagger})$ with $E(t)^2<0$. 
This implies that the instability is present:
From the equation of motion $-\partial_t^2\Phi_i=E(t)^2\Phi_i$ with $E(t)^2<0$,
we can see that the field $\Phi_i$ increases endlessly because
a particle with its coordinate $\Phi_i$ rolls down the slope given by $E(t)^2|\Phi_i|^2<0$.

\vspace{0.2cm}

We have found that even if $gB$ varies spatially or temporally,
the instability arises owing to the potential term ( $-2gB$ ). The growth rate of the unstable modes
is, however, much smaller than that in the case of homogeneous $gB$. The small growth rate can be
represented 
by using effective homogeneous magnetic field $B_{\rm{eff}}$; $\gamma=\sqrt{gB_{\rm{eff}}}$.
This effective gauge field gives rise to the imaginary ``mass" $\sqrt{-2gB_{\rm{eff}}}$
of the Nielsen-Olesen unstable modes so that
the modes increase exponentially.
The fact can be described by the following effective Lagrangian of the 
Nielsen-Olesen unstable modes $\phi_{NO}\equiv (\Phi_1+i\Phi_2)/\sqrt{2}$,

\begin{equation}
\label{NO}
L_{\rm{eff}}=|(\partial_{\mu}+igA_{\rm{eff},\mu})\phi_{NO}|^2+2gB_{\rm{eff}}|\phi_{NO}|^2
=|\partial_t\phi_{NO}|^2-|(\partial_3+igA_{\rm{eff},3})\phi_{NO}|^2
+gB_{\rm{eff}}|\phi_{NO}|^2,
\end{equation} 
where homogeneous background magnetic field is described by
$B_{\rm{eff}}=\partial_1A_{\rm{eff},2}-\partial_2A_{\rm{eff},2}$. 
The numerical simulations of non-expanding glasma
show that $\sqrt{gB_{\rm{eff}}}$ is about $(1\sim 2)\times 10^{-1}Q_s$.
Furthermore, we assumed that
the effect of the inhomogeneous electric field is also described by using
the effective homogeneous electric field, $E_{\rm{eff}}=\partial_0 A_{\rm{eff},3}$.
  

\section{monopole production and rapid decay of glasma}
\label{3}

It is generally believed that
magnetic monopoles in QCD play the role in confining 
quarks. They condense in vacuum and form dual superconductors
in which color electric flux is squeezed.
In lattice gauge theories, 
we can see such a role of the magnetic monopoles with the use of maximal Abelian gauge.
Furthermore,
effective theories of dual superconductors have been
explored\cite{koma} in which magnetic monopoles are described by a complex scalar
field. In the theories we have dual gauge fields $A^d_i$
which couple minimally with the monopole field.
Electric and magnetic fields are described such that $E_i=\epsilon_{i,j,k}\partial_j A^d_k$
and $B_i=-\partial_0 A^d_i-\partial_i A^d_0$, respectively. 
We assume an effective model of the dual superconductor used to see
the effects of the monopoles in high energy heavy ion collisions.
( The roles of their effects in thermalized QGP have also been discussed; for instance,
small viscosity of QGP would be caused by the monopoles\cite{vis}. )

Now, we consider the production of magnetic monopoles and their back reaction to the background 
magnetic field by using classical statistical field theory. 
These monopoles are produced by
the color magnetic fields owing to Schwinger mechanism\cite{sch}. 

Effective Lagrangian of the monopole field $\phi$ are given by

\begin{equation}
\label{MO}
L=|D_{\mu}^d\phi|^2+m^2|\phi|^2-\frac{\lambda}{4}|\phi|^4
\end{equation}
with $D_{\mu}^d=\partial_{\mu}+ig_{\rm{m}}A_{\mu}^d$,
where $g_{\rm{m}}$ denotes magnetic charge ( $=2\pi/g$ ) of the monopoles.
Phenomenologically the parameter $m$ approximately takes a value of the order of $1$GeV \cite{koma}.
Obviously, the mass is homogeneous 
( independent of space-time ).
In this section we consider the effective homogeneous 
background gauge fields $B_{\rm{eff}}$ and $E_{\rm{eff}}$
mentioned above.

Here, we note a similarity between the monopoles in eq(\ref{MO}) and the Nielsen-Olesen unstable modes
in eq({\ref{NO}).
They occupy the Landau levels under the electric ( magnetic ) field.
The monopoles possess the imaginary mass $\sqrt{-m^2}$, while the unstable modes
possess the imaginary ``mass" $\sqrt{-2gB_{\rm{eff}}}$.
They are produced by Schwinger mechanism; monopoles ( Nielsen-Olesen unstable modes ) 
under magnetic ( electric ) field. 
Both of them are described by the scalar fields; more precisely, scalar field in the monopoles
and scalar-like field in the unstable modes 
( only the component of the vector field with the magnetic moment parallel to $\vec{B}$ ). 

\vspace{0.2cm}
First, by
using the standard formulae\cite{tanji,ita} of the Schwinger mechanism,
we compare the production rate of the monopoles with that of the Nielsen-Olesen unstable modes.
The production rate is proportional to the number of the particles produced
in unit volume.
The production rate of the monopole under the magnetic field $B_{\rm{eff}}$
is given by
$\exp\big(\pi (m^2-g_{\rm{m}}E_{\rm{eff}})/g_{\rm{m}}B_{\rm{eff}}\big)$, 
where we assumed that the monopoles occupy the lowest Landau level.
On the other hand,
the production rate\cite{ita} of Nielsen-Olesen unstable modes
is given by $\exp(\pi gB_{\rm{eff}}/gE_{\rm{eff}})$.
( The energy of the lowest Landau level is given by 
$\sqrt{p_3^2+g_mE_{\rm{eff}}-m^2}$ for the monopoles, while $\sqrt{p_3^2-gB_{\rm{eff}}}$
for the Nielsen-Olesen unstable modes. Hence
the production rate of the monopoles can be obtained 
in the similar way to the one\cite{ita} in Nielsen-Olesen unstable modes.)

\vspace{0.2cm}
Hereafter, for definiteness, we use the parameters $m=0.7$ GeV, $Q_s=2$ GeV, $\alpha_s(Q_s)=1/3$ 
and $\sqrt{gB_{\rm{eff}}}=0.17\,Q_s$
( the value $\sqrt{gB_{\rm{eff}}}$ has been estimated using the results in the numerical
simulation\cite{berges4} ). Then,
$g_mB_{\rm{eff}}=(2\alpha_s)^{-1}gB_{\rm{eff}}=(0.42\rm{GeV})^2$ with $\alpha_s\equiv g^2/4\pi$.

\vspace{0.2cm}
We assume that $g_mE_{\rm{eff}}\sim g_mB_{\rm{eff}}$ which holds just after the heavy ion collisions.
Then, we find that the production rate  
$r(m)\equiv\exp\big(\pi (m^2/g_{\rm{m}}B_{\rm{eff}}-1)\big)\simeq\exp(1.8\pi)$
of the monopoles is about 10 times larger than the rate $r(N)\equiv \exp(\pi)$ of
Nielsen-Olesen unstable modes; $r(m)/r(N)=\exp(0.8\pi)\simeq 12$. This means that
the number of the monopoles produced is $10$ times larger than the number of Nielsen-Olesen
unstable modes.  In the estimation,
the back reaction of the particle production to the background gauge fields
is not taken into account.



\vspace{0.3cm}
Now, we consider the back reaction of the monopole production by using classical statistical
field theory\cite{csft2}. We assume
homogeneous color magnetic $\vec{B}=(0,0,B_{\rm{eff}})$ 
and electric fields $\vec{E}=(0,0,E_{\rm{eff}})$.
Under the background electric field, the states of the magnetic monopoles
are characterized by Landau levels. Among them, we only consider the production of the monopoles
in the lowest Landau level whose wavefunctions are given by

\begin{equation}
\phi\equiv (x_1-ix_2)^n\exp(-\frac{g_mE_{\rm{eff}}|z|^2}{4}+ip_3x_3),
\end{equation}
with $z\equiv x_1+ix_2$ and integer $n\ge 0$
where we used a gauge potential $\vec{A}_d=(-E_{\rm{eff}}x_2/2,E_{\rm{eff}}x_1/2,0)$.
The states are localized in transverse space.
But, by taking the appropriate linear combination of the wavefunctions we
can form almost homogeneous field configurations
in the transverse plane.
Then, their magnetic currents $J_m(\phi)$ are also almost homogeneous so that the effects of
the back reaction ( $\partial_t B_{\rm{eff}}=-J_m$ ) to the monopole production is
homogeneous.
Such field configurations are given by,

\begin{equation}
\label{sum}
\phi=\sum_{l=1\sim N}\psi_l(\vec{x}), \quad \psi_l(\vec{x})
=\int dp_3 \,c(p_3)\exp(-\frac{g_mE_{\rm{eff}}|z-z_l|^2}{4}+ip_3x_3),
\end{equation} 
where $c(p_3)$ is a dimensionless function of the longitudinal momentum $p_3$ 
and $z_l\equiv x_{1.l}+ix_{2,l}$.
Each component $\psi_l$ satisfies the condition, 
$\psi_l\psi_{l'}\simeq \delta_{l,l'}\psi_l^2$ 
because we impose that $|z_l-z_{l'}|\geq l_E\equiv \frac{1}{\sqrt{g_mE_{\rm{eff}}}}$.
Namely, a configuration $\psi_l$ is separated with the nearest neighbors 
approximately by the distance larger than $l_E$.
Furthermore, we assume that the area $L^2$ 
of the transverse plane is given by $L^2=k\,Nl^2_E$ where $k$ represents
how dense the transverse plane is occupied by the field $\psi_l$.
( Their number density is given by $N/L^2\propto 1/k$. )
In order for our approximation to hold, we should take $k$ such that $k$ is not too
small ( $k<1$ ) to avoid over dense configuration and not too large ( $k>100$ ) to avoid inhomogeneity
in the transverse plane.
For definiteness, we assume $k=10$ so that each field configuration $\psi_l$ is separated from each others
by the distance equal to or larger than $\simeq 3l_E$. 
Then, the field configuration $\phi$ is 
approximately homogeneous.
This kind of the field configuration was analyzed\cite{ninomiya} to discuss so called
``spaghetti vacuum". 

Using the field configuration, we write down the energies of the monopoles and
the magnetic field,

\begin{eqnarray}
H_m&=&\int d^3x \Bigg(\frac{1}{2}(\partial_tA_d)^2+|\partial_t\phi|^2+|D_i^d\phi|^2
-m^2|\phi|^2\Bigg) \nonumber \\
&\simeq & \int d^3x 
\Bigg(\frac{1}{2}(\partial_tA_d)^2+N(|\partial_t\psi|^2+|(i\partial_3-g_mA_d)\psi|^2
+(g_mE_{\rm{eff}}-m^2)|\psi|^2)\Bigg) \nonumber \\
&=&N\int dx_3 \Bigg(\frac{kl_E^2}{2}(\partial_tA_d)^2+(\partial_tC)^2+|(\partial_3+ig_mA_d)C|^2
+(g_mE_{\rm{eff}}-m^2)|C|^2\Bigg) \nonumber \\
&=&NL\frac{kl_E^2}{2}(\partial_tA_d)^2+N\int dp \Bigg(|\partial_t C_p|^2+|(p+g_mA_d)C_p|^2+
(g_mE_{\rm{eff}}-m^2)|C_p|^2\Bigg)
\end{eqnarray}
with $\int dx_3=L$,  
where 
\begin{equation}
\psi\equiv C(x_3,t)\exp(-g_mE_{\rm{eff}}|z|^2/4)\sqrt{\frac{g_mE_{\rm{eff}}}{2\pi}}=
\int dp \,\frac{C_p}{\sqrt{2\pi}}\,\exp(ipx_3-g_mE_{\rm{eff}}|z|^2/4)\sqrt{\frac{g_mE_{\rm{eff}}}{2\pi}}.
\end{equation}

The color magnetic field is given by $B_{\rm{eff}}=\partial_0A_d$ with 
the homogeneous dual gauge potential $A_d\equiv A_{3,d}$.
Here, we assumed that the gauge potential $A_d$ is homogeneous both in the transverse and 
longitudinal directions. That is, we wish to study how the homogeneous component of 
the magnetic field produced just after heavy ion collisions decreases 
by the monopole production.
Furthermore, we assume that
the quartic term $|\phi|^4$ is negligibly small. 

Using the Hamiltonian, we can derive the equation of motions of the fields $C_p(t)$,

\begin{equation}
\label{eqm}
\partial_t^2C_p=(m^2-g_mE_{\rm{eff}})C_p-(p+g_mA_d)^2C_p \quad \mbox{and} \quad 
L\partial_t^2(g_mA_d)l_{\rm{E}}^2=-\frac{2g_m^2}{k\pi}\int_{-\infty}^{+\infty} dp(p+g_mA_d)|C_p|^2,
\end{equation}
with $k=10$,
where we have written down a dual Maxwell equation of the gauge field $\partial_t B=\partial_0^2A_d=-J_m$
with the magnetic current $J_m\equiv\frac{2g_m}{Ll_{\rm{E}}^2k\pi}\int_{-\infty}^{+\infty} dp(p+g_mA_d)|C_p|^2$.
It describes how the magnetic field decreases by the effect of the magnetic current $J_m$,
which is induced by the monopole production.

To solve the equation, we need to impose initial conditions of
$A_d$ and $C_p$. The initial condition of $A_d$ is given by
$A_d(t=0)=0$ and $B_{\rm{eff}}(t=0)=\partial_tA_d(t=0)=B_0$ where $B_0$ is the initial value of
the magnetic field; $g_mB_0=g_mE_{\rm{eff}}=(0.42\mbox{GeV})^2 \ll Q_s^2=(2\mbox{GeV})^2$.
On the other hand, we take initial conditions of the monopole field, 
following the discussion by Dusling et al.\cite{csft}. By using the initial conditions, we can take into account
one loop quantum effects of the monopoles in our classical calculation.
The use of the initial conditions is the essence of the classical statistical field theory.

The initial conditions are given in the following,

\begin{eqnarray}
C_p(t=0)&=&\frac{\exp(-\pi(1-\bar{m}^2)/8)}{(2g_mE_{\rm{eff}})^{1/4}}\Big(D_{(-1+i(\bar{m}^2-1))/2}
\Big(\frac{\sqrt{2}pe^{i\pi/4}}{\sqrt{g_mE_{\rm{eff}}}}\Big)d_p+
\bar{D}_{(-1+i(\bar{m}^2-1))/2}\Big(\frac{-\sqrt{2}pe^{i\pi/4}}{\sqrt{g_mE_{\rm{eff}}}}\Big)f_p\Big) \nonumber \\
\partial_tC_p(t=0)&=&\frac{\exp(-\pi(1-\bar{m}^2)/8)}{(2g_mE_{\rm{eff}})^{1/4}}\partial_t\Big(
D_{(-1+i(\bar{m}^2-1))/2}
\Big(\frac{\sqrt{2}(t\,g_mE_{\rm{eff}}+p)e^{i\pi/4}}{\sqrt{g_mE_{\rm{eff}}}}\Big)d_p +\nonumber \\
&+&\bar{D}_{(-1+i(\bar{m}^2-1))/2}
\Big(\frac{\sqrt{2}(t\,g_mE_{\rm{eff}}-p)e^{i\pi/4}}{\sqrt{g_mE_{\rm{eff}}}}\Big)f_p\Big) \quad \mbox{as} \quad t\to 0
\end{eqnarray}
with parabolic cylinder function $D_{\nu}(z)$ and $\bar{m}^2\equiv m^2/g_mE\simeq 1.68$, 
where $d_p$ and $f_p$ denote Gaussian random variables satisfying

\begin{equation}
\label{random}
\langle d_p\bar{d_q}\rangle=\langle f_p\bar{f_q}\rangle=\delta(p-q), \quad 
\langle d_pd_q\rangle=\langle f_pf_q\rangle=\langle d_p\bar{f_q}\rangle=\langle f_p\bar{d_q}\rangle=
\langle \bar{d_p}\bar{d_q}\rangle=\langle \bar{f_p}\bar{f_q}\rangle=0. 
\end{equation}

The average $\langle\,\,\rangle$ over the Gaussian random variables is taken 
after solving the equations (\ref{eqm}). We should note that the average $\langle |C_p|^2\rangle $
is proportional to $2\pi\delta(p=0)=\int dx_3 \exp(ix_3p)_{p=0}=L$. 
Hence, the factor $L$ in the denominator of $J_m$
is cancelled with that of $\langle |C_p|^2\rangle $ so that the factor $L$ 
disappears in eq(\ref{eqm}).

By solving these equations with the initial conditions and $E_{\rm{eff}}=B_{\rm{eff}}(t=0)$, 
we can find how fast the magnetic field 
$B_{\rm{eff}}=\partial_tA_d$ decays.
The decay is caused by the production of the magnetic monopoles.

For comparison, we write down the corresponding equations for Nielsen-Olesen unstable modes,
which describe the decay of electric field caused by the production of the unstable modes.
The equations are in the following,

\begin{equation}
\label{eqmN}
\partial_t^2C_p^N=gB_{\rm{eff}}C_p^N-(p+gA)^2C_p^N \quad \mbox{and} \quad 
L\partial_t^2(gA)l_{\rm{B}}^2=-\frac{2g^2}{k\pi}\int_{-\infty}^{+\infty} dp(p+gA)|C_p^N|^2,
\end{equation}
with $l_{\rm{B}}\equiv \sqrt{1/gB_{\rm{eff}}}$ and the initial conditions,

\begin{eqnarray}
\label{randomN}
C_p^N(t=0)&=&\frac{\exp(\pi/8)}{(2gB_{\rm{eff}})^{1/4}}\Big(D_{(-1+i)/2}
\Big(\frac{\sqrt{2}pe^{i\pi/4}}{\sqrt{gB_{\rm{eff}}}}\Big)d_p+
\bar{D}_{(-1+i)/2}\Big(\frac{-\sqrt{2}pe^{i\pi/4}}{\sqrt{gB_{\rm{eff}}}}\Big)f_p\Big) \nonumber \\
\partial_tC_p^N(t=0)&=&\frac{\exp(\pi/8)}{(2gB_{\rm{eff}})^{1/4}}\partial_t\Big(
D_{(-1+i)/2}
\Big(\frac{\sqrt{2}(t\,gB_{\rm{eff}}+p)e^{i\pi/4}}{\sqrt{gB_{\rm{eff}}}}\Big)d_p +\nonumber \\
&+&\bar{D}_{(-1+i)/2}
\Big(\frac{\sqrt{2}(t\,gB_{\rm{eff}}-p)e^{i\pi/4}}{\sqrt{gB_{\rm{eff}}}}\Big)f_p\Big) \quad \mbox{as} \quad t\to 0
\end{eqnarray}
with $A(t=0)=0$ and $\partial_tA(t=0)=E_{\rm{eff}}$,
where $d_p$ and $f_p$ are the random variables satisfying the above equations (\ref{random}).
( In the previous paper\cite{iwazaki} we have used initial conditions for Nielsen-Olesen unstable modes
different from the ones used in the present paper. Using the initial conditions in the present paper,
we can appropriately take into account of quantum effects of the unstable modes on the background electric field. )

\vspace{0.2cm}
Before solving the above equations, we note that the average $\langle \,|C_p|^2\rangle $
is given such that

\begin{equation}
\langle \,|C_p(t=0)|^2\,\rangle= \frac{L\exp(-\pi(1-\bar{m}^2)/4)}{2\pi(2g_mE_{\rm{eff}})^{1/2}}\Big(|D_{(-1+i(\bar{m}^2-1))/2}
\Big(\frac{\sqrt{2}pe^{i\pi/4}}{\sqrt{g_mE_{\rm{eff}}}}\Big)|^2+
|\bar{D}_{(-1+i(\bar{m}^2-1))/2}\Big(\frac{-\sqrt{2}pe^{i\pi/4}}{\sqrt{g_mE_{\rm{eff}}}}\Big)|^2\Big).
\end{equation}
Then, instead of taking the average over the random variables $d_p$ and $f_p$
after solving the equations, for simplicity, we use an initial condition,

\begin{eqnarray}
\tilde{C}_p(t=0)&=&\frac{\exp(-\pi(1-\bar{m}^2)/8)}{(2g_mE_{\rm{eff}})^{1/4}}\Big(|D_{(-1+i(\bar{m}^2-1))/2}
\Big(\frac{\sqrt{2}pe^{i\pi/4}}{\sqrt{g_mE_{\rm{eff}}}}\Big)|^2+
|\bar{D}_{(-1+i(\bar{m}^2-1))/2}\Big(\frac{-\sqrt{2}pe^{i\pi/4}}{\sqrt{g_mE_{\rm{eff}}}}\Big)|^2\Big)^{1/2}
\nonumber \\
\partial_t \tilde{C}_p(t=0)&=&\frac{\exp(-\pi(1-\bar{m}^2)/8)}{(2g_mE_{\rm{eff}})^{1/4}}\partial_t
\Big(|D_{(-1+i(\bar{m}^2-1))/2}\Big(\frac{\sqrt{2}(tg_mE_{\rm{eff}}+p)e^{i\pi/4}}{\sqrt{g_mE_{\rm{eff}}}}\Big)|^2 \nonumber \\
&+&
|\bar{D}_{(-1+i(\bar{m}^2-1))/2}\Big(\frac{\sqrt{2}(tg_mE_{\rm{eff}}-p)e^{i\pi/4}}{\sqrt{g_mE_{\rm{eff}}}}\Big)|^2\Big)^{1/2}
\quad \mbox{as} \quad t\to 0,
\end{eqnarray}
where we have rewritten $C_p$ such that $C_p=\sqrt{L/2\pi}\,\tilde{C}_p$.

In Fig.1 we show how the background color electric field decays by the production of
the Nielsen-Olesen unstable modes. Similarly, in Fig.2 we show 
how the background color magnetic field decays by the production of
the magnetic monopoles.
We can see that the decay of the magnetic field proceeds very rapidly, about 10
times more rapidly than the decay of the electric field.
The difference in the life times of the gauge fields comes from the difference 
in the initial conditions. Namely, the
initial amplitude $C_p(t=0)$ of the monopole field is much larger than
that of the Nielsen-Olesen unstable modes $C_p^N(t=0)$
( Fig.3 ). 
Physically, it means that
the number of the monopoles produced spontaneously at $t=0$ is very large 
compared with the number of the Nielsen-Olesen unstable modes.
The difference is consistent with the difference in the production rate of
the monopoles and Nielsen-Olesen unstable modes by Schwinger mechanism;
the production rate of the monopoles is 10 times larger than that of Nielsen-Olesen
unstable modes.
Although the field amplitude $C_{p=0}(t)$ begins to grow such as $C_{p=0}\propto \exp(tm)$
for $t>m^{-1}\simeq 0.3 \rm{fm/c}$,
the magnetic field decays before the start of the growth. Thus, the exponential
growth of the amplitude does not contribute to the decay of the magnetic field.
It decays mainly due to the large amount of the spontaneous monopole production. 

\begin{figure}[htb]
  \centering
  \includegraphics*[width=65mm]{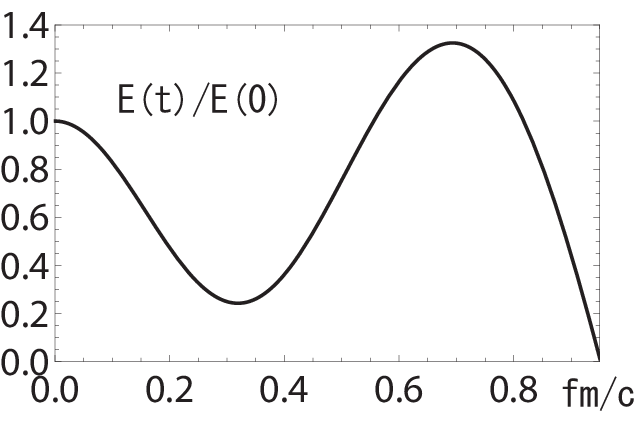}
  \caption{The decay of the color electric field by
the production of the Nielsen-Olesen unstable modes.}
     \label{f1}
\end{figure}  
  
\begin{figure}[htb]
  \centering
  \includegraphics*[width=65mm]{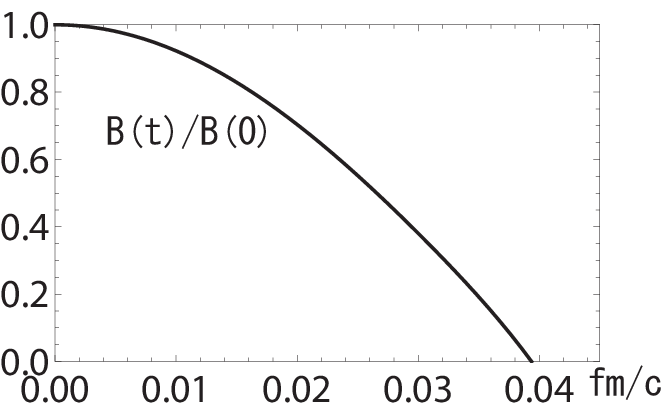}
  \caption{The decay of the color magnetic field by the production
of the magnetic monopoles.}
     \label{f2}
\end{figure}

\begin{figure}[htb]
  \centering
  \includegraphics*[width=65mm]{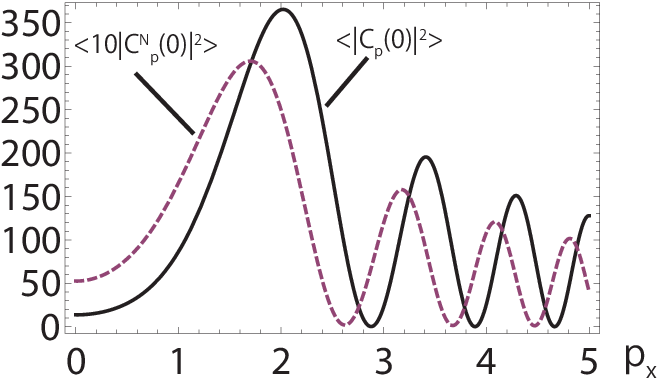}
  \caption{$p_x\equiv p/\sqrt{gB_{\rm{eff}}}$ in $10\times\langle |C_p^N|^2\rangle$ 
and $p_x\equiv p/\sqrt{g_mE_{\rm{eff}}}$ 
in $\langle |C_p|^2\rangle$}
     \label{f3}
\end{figure}

Here we make a comment on the magnetic current $J_m$ in eq(\ref{eqm}).
It vanishes at $t=0$ because the integrand is antisymmetric in the variable $p$;
$C_p(t=0)=C_{-p}(t=0)$.
That is, the monopole current is absent at $t=0$.
Then, the current begins to flow after the spontaneous production of the magnetic monopoles.
The strength of the current is determined by the amplitude of the monopoles $C_p$.
This initial large amount of the monopole production causes the magnetic field decay rapidly.

\section{discussions and conclusions}
\label{4}

Classical solutions of color magnetic monopoles
are unstable unless their stability is guaranteed topologically.
Indeed, there are no solutions of stable magnetic monopoles in real QCD.
It has been shown\cite{coleman} that growth rates of unstable modes around 
classical monopoles
can be infinitely large; the rates are proportional to the logarithm of the volume in 
the system. The fact indicates that the monopoles rapidly decay into gluons
after their production. Furthermore, they couple strongly with gluons
because the smaller the $g$, the larger the coupling $g_m=2\pi/g$. Therefore,
thermalized QGP would be generated immediately
after the decay of the background gauge fields.
 
Background gauge fields used in previous numerical simulations does not involve 
such classical monopoles,
because the background gauge fields
are homogeneous in the longitudinal direction. ( The gauge fields produced by monopoles
are not homogeneous. )
Thus, only unstable modes with finite growth rates arise in the simulations. 
If the monopoles are appropriately taken into account in numerical simulations,
unstable modes with infinitely large growth rates
would be found. 

In our calculation background electric fields have been supposed to be 
static. But, real
background electric fields oscillate with time so that magnetic fields are newly generated.
Such magnetic fields would rapidly decay by the production of the monopoles. 
Therefore, the electric fields also decay.
That is, the gauge fields in the glasma decays by the monopole production.

\vspace{0.2cm}
We have shown by using classical statistical field theory 
that the background color magnetic fields of glasma rapidly decay by the production of 
the color magnetic monopoles. The life time $t_c$ is sufficiently short so that
the phenomenological constraint ( $t_c < 1$fm/c ) is satisfied.
Although there are several ambiguous numerical factors in our calculation,
it shows that the monopoles play important roles in the rapid decay of the glasma.




\end{document}